\documentclass[twocolumn,aps,showpacs,amsmath,amsfonts,amssymb,floatfix,pre]{revtex4}
\usepackage{graphicx}
\usepackage{dcolumn}
\usepackage{bm}

\begin{document}

\title{Random matrix analysis of complex networks}

\author{Sarika Jalan} 
\email{sarika@mpipks-dresden.mpg.de}
\author{Jayendra N. Bandyopadhyay} 
\email{jayendra@mpipks-dresden.mpg.de}
\affiliation{Max-Planck Institute for the 
Physics of Complex Systems, N\"{o}thnitzerstr. 38, D-01187 Dresden, Germany}

\begin{abstract} 

We study complex networks under random matrix theory (RMT) framework. Using 
nearest-neighbor and next-nearest-neighbor spacing distributions we analyze the 
eigenvalues of adjacency matrix of various model networks, namely, random, 
scale-free and small-world networks. These distributions follow Gaussian orthogonal 
ensemble statistic of RMT. To probe long-range correlations in the eigenvalues we 
study spectral rigidity via $\Delta_3$ statistic of RMT as well. It follows RMT 
prediction of linear behavior in semi-logarithmic scale with slope being $\sim 
1/\pi^2$. Random and scale-free networks follow RMT prediction for very large scale. 
Small-world network follows it for sufficiently large scale, but much less than the 
random and scale-free networks.

\end{abstract}

\maketitle

\section{Introduction} 
\label{intro} 

Random matrix theory (RMT), initially proposed to explain statistical properties of 
nuclear spectra, had successful predictions for the spectral properties of different 
complex systems such as disordered systems, quantum chaotic systems, large complex 
atoms, etc., followed by numerical and experimental verifications in the last few 
decades \cite{mehta,rev-rmt}. Quantum graphs, which model the systems of interest in 
quantum chemistry, solid state physics and transmission of waves, have also been 
studied under the RMT framework \cite{Qgraph}. Recently, RMT has been shown to be 
useful also in understanding the statistical properties of empirical 
cross-correlation matrices appearing in the study of multivariate time series of 
followings: price fluctuations in stock market \cite{rmt-stock1,rmt-stock2}, EEG 
data of brain \cite{rmt-brain}, variation of different atmospheric parameters 
\cite{rmt-atmosphere}, etc.

In our previous studies \cite{pap1,pap2} complex networks have been analyzed under 
RMT framework.  These works consider nearest-neighbor spacing distribution (NNSD) of 
eigenvalues spectra of adjacency and Laplacian matrices of various extensively 
studied networks. The NNSD gives probability for finding neighboring eigenvalues 
with a given spacing, and it follows two universal properties depending upon 
underlying correlations among the eigenvalues. For the correlated eigenvalues, NNSD 
follows Gaussian orthogonal ensemble (GOE) statistics of RMT, whereas it follows 
Poissonian statistics for the uncorrelated eigenvalues. One of the main advantages 
of RMT approach is that depending on the nature of eigenvalues correlations one can 
separate system dependent part from random universal part, which are intermingled 
due to the complexity of the system 
\cite{rev-rmt,rmt-stock1,rmt-stock2,rmt-brain,rmt-atmosphere}. RMT analysis for the 
various networks shows that the NNSD of complex networks also follow universal GOE 
statistics of RMT \cite{pap1}. This finding suggests that different results of GOE 
statistics, which have successfully been applied to understand the systems coming 
from various fields starting from nuclei to the stock-market, can be applied to 
study networks as well. Our earlier works concentrate on the NNSD studies of 
networks. NNSD carries information for the correlation between two adjacent 
eigenvalues, but do not tell about the correlation between two far-off eigenvalues. 
Therefore, even though NNSD follows GOE statistics of RMT, other properties may show 
deviations, which suggests that one can not rely on NNSD results exclusively.  To 
probe for long-range correlations as well, current work considers spectral rigidity 
test via well known $\Delta_3$-statistic of RMT. It is found that the spectral 
rigidity of the complex networks follows RMT prediction, with scale depending upon 
the properties of the networks. Present work also analyzes the {\it 
next}-nearest-neighbor spacing distribution (nNNSD) of the adjacency matrix of the 
networks.

The paper is organized as follows: following the above introduction, Sec. 
\ref{network} explains various aspects of complex networks studies. Sec. \ref{RMT} 
describes some basics of RMT relevant to our studies. Sec. \ref{results} illustrates 
the RMT analysis for various model networks, namely; random, scale-free and small 
world. The NNSD is the most widely studied property in random matrix literature, 
therefore, this section includes NNSD results for the above mentioned model networks 
\cite{pap1}, and presents results for nNNSD and the $\Delta_3$-statistic of these 
networks. Finally, Sec.  \ref{summary} discusses and summarizes results with some 
possible future directions.

\section{Complex networks}
\label{network}

Last 10 years have witnessed a rapid advancement in the studies of complex networks. 
The main concept of the network theory is to define complex systems in terms of 
networks of many interacting units. Few examples of such systems are interacting 
molecules in living cell, nerve cells in brain, computers in Internet communication, 
social networks of interacting people, airport networks with flight connections, etc 
\cite{rev-Strogatz,rev-network,rev-Boccaletti}. In the graph theoretical 
terminology, units are called nodes and interactions are called edges \cite{graph}. 
Various model networks have been introduced to study the behavior of complex systems 
having underlying network structures. These model networks are based on simple 
principles, still they capture essential features of the underlying systems.

\subsection{Structural properties}

In random graph model of Erd\"os and R\'enyi any two nodes are randomly connected 
with probability $p$ \cite{erdos}. This model assumes that interactions between 
nodes are random. Recently, with the availability of large maps of real world 
networks, it has been observed that the random graph model is not appropriate for 
studying the behavior of real world networks. Hence many new models have been 
introduced. Watts and Strogatz proposed a model, popularly known as `small-world 
network', which has properties of small diameter and high clustering \cite{SW}. 
Moreover, this model network is very sparse : network with a very few number of 
edges, another property shown by many real-world networks. In addition to the above 
mentioned properties, Barab\'asi and Albert show that degree distributions of many 
real-world networks have power-law. This implies that some nodes are much more 
connected than the others \cite{BA}. Barab\'asi-Albert's scale-free model and 
Watts-Strogatz's small-world model have contributed immensely in understanding 
evolution and behavior of the real systems having network structures. Following 
these two new models came an outbreak in the field of networks. These studies show 
that real world networks have coexistence of {\it randomness} and {\it regularity} 
\cite{rev-Strogatz,Newman,rev-Costa}.

\subsection{Spectral properties} 

Apart from the above mentioned investigations which focus on direct 
measurements of the structural properties of networks, there have been lot of studies 
demonstrating that properties of networks or graphs could be 
well characterized by the spectrum of associated adjacency matrix $A$ \cite{spectrum}. 
For an unweighted graph, it is defined in the following way
: $A_{ij} = 1$, if $i$ and $j$ nodes are connected and $zero$ otherwise. For an undirected 
network, this matrix is symmetric and consequently has real eigenvalues. Eigenvalues 
give information about some basic topological properties of the underlying network 
\cite{spectrum,handbook}. Spectral properties of networks have also been used to 
understand some of the dynamical properties of interacting chaotic units on 
networks, for example largest eigenvalue of adjacency matrix determines transition 
to the synchronized state \cite{ott}. The distribution of the eigenvalues of a 
matrix having finite probability $p$ of nonzero Gaussian distributed random elements 
per row, follows Wigner semicircular law in the limit $p \rightarrow 1$. For very 
small $p$ also, which corresponds to the sparse random matrix, one gets semicircular 
law with several peaks at different eigenvalues \cite{sparseRM}.

With the increasing availability of large maps of real-world networks, analyses of 
spectral densities of adjacency matrix of real-world networks and model networks 
having real-world properties have also begun \cite{Vicsek,Dorogovtsev,Kim,Aguiar}. 
These analyses show that the matrix constructed by $zero$ and $one$ elements 
corresponding to a unweighted random network, also follows Wigner semicircular law 
\cite{Vicsek} with degeneracy at $\lambda=0$. Small-world model networks show very 
complex spectral density with many sharp peaks \cite{Aguiar}, while the spectral 
density of scale-free model networks exhibits so called triangular distribution 
\cite{Vicsek,Kim,Aguiar}. Spectral density and NNSD of the random matrices 
constructed by $zero$ and $one$ elements
have been studied extensively in the Ref. \cite{bauer}. These studies show that
NNSD of the random matrices follow GOE distribution of RMT.

\section{Random matrix statistics} 
\label{RMT}

In the random matrix studies of eigenvalues spectra, one has to consider two kinds 
of properties : (1) global properties, like spectral density or distribution of 
eigenvalues $\rho(\lambda)$, and (2) local properties, like eigenvalue fluctuations 
around $\rho(\lambda)$. Among these, the eigenvalue fluctuations is the most popular 
one. This is generally obtained from the NNSD of the eigenvalues. The eigenvalues of 
network are denoted by $\lambda_i,\,\,i=1,\dots,N$, where $N$ is the size of the 
network and $\lambda_i < \lambda_{i+1}, \, \, \forall i$. In order to get universal 
properties of the eigenvalue fluctuations, one has to remove the spurious effects 
due to the variations of spectral density and to work at constant spectral density 
on the average. Thereby, it is customary in RMT to unfold the eigenvalues by a 
transformation $\overline{\lambda}_i = \overline{N} (\lambda_i)$, where 
$\overline{N} (\lambda) = \int_{\lambda_{\mbox{\tiny min}}}^\lambda\, 
\rho(\lambda^\prime)\, d \lambda^\prime$ is the averaged integrated eigenvalue 
density \cite{mehta}. Since analytical form for $\overline{N}$ is not known, we 
numerically unfold the spectrum by polynomial curve fitting. Using the unfolded 
spectrum, we calculate the nearest-neighbor spacings as
\begin{equation} 
s_1^{(i)}=\overline{\lambda}_{i+1}-\overline{\lambda}_i;
\nonumber
\end{equation}
and due to the above unfolding, the average nearest-neighbor spacings $\langle s_1 
\rangle$ becomes {\it unity}, being independent of the system. The NNSD $P(s_1)$ is 
defined as the probability distribution of these $s_1^{(i)}$'s. In case of Poisson 
statistics,
\begin{equation}
P(s_1)=\exp(-s_1); 
\end{equation}
whereas for GOE 
\begin{equation}
P(s_1)=\frac{\pi}{2} s_1\exp \left(-\frac{\pi s_1^2}{4}\right). 
\label{goe}
\end{equation}
For the intermediate cases, NNSD is described by Brody formula \cite{brody}: 
\begin{subequations}
\label{eq-brody}
\begin{equation}
P_{\beta}(s_1) = A s_1^\beta\exp\left(-B s_1^{\beta+1}\right),
\end{equation}
where $A$ and $B$ are determined by the parameter $\beta$ as follow :
\begin{equation}
A\,=\,(1+\beta) \alpha\,\mbox{and}\,\alpha = \left[\Gamma\left(\frac{\beta+2}
{\beta+1}\right)\right]^{\beta+1}.
\end{equation} 
\end{subequations} 
This is a semiempirical formula characterized by the single parameter $\beta$, 
popularly known as Brody parameter. $\beta=1$ corresponds to the GOE statistics and 
$\beta=0$ corresponds to the Poisson statistics.

Apart from NNSD, the {\it next}-nearest-neighbor spacings distribution (nNNSD) is 
also used to characterize the statistics of eigenvalues fluctuations. We calculate 
this distribution $P(s_2)$ of next-nearest-neighbor spacings
\begin{equation}
s_2^{(i)} = (\overline{\lambda}_{i+2} - \overline{\lambda}_i)/2 
\label{eq-s2}
\end{equation}
between the unfolded eigenvalues. Factor of {\it two} at the denominator is inserted 
to make the average of next-nearest-neighbor spacings $\langle s_2 \rangle$ {\it 
unity}. According to Ref. \cite{mehta}, the nNNSD of GOE matrices is identical to 
the NNSD of Gaussian symplectic ensemble (GSE) matrices, i.e.,
\begin{equation}
P(s_2) = \frac{2^{18}}{3^6 \pi^3}\,s_2^4 \exp\left(-\frac{64}{9 \pi}\,s_2^2
\right).
\label{spacing_GSE}
\end{equation}
The NNSD and nNNSD reflect only local correlations among the eigenvalues. The 
spectral rigidity, measured by the $\Delta_3$-statistic of RMT, gives information 
about the long-range correlations among eigenvalues and is more sensitive test for 
RMT properties of the matrix under investigation \cite{mehta,casati}. In the 
following we describe the procedure to calculate this quantity.
 
The $\Delta_3$-statistic measures the least-square deviation of the spectral 
staircase function representing the averaged integrated eigenvalue density 
$\overline{N}(\lambda)$ from the best straight line fitting for a finite interval 
$L$ of the spectrum, i.e.,
\begin{equation}
\Delta_3(L; x) = \frac{1}{L} \min_{a,b} \int_x^{x+L} \,\left[ 
N(\overline{\lambda}) - a \overline{\lambda} -b \right]^2\,d \overline{\lambda}
\label{eq-delta3}
\end{equation}
where $a$ and $b$ are obtained from a least-square fit. Average over several choices 
of $x$ gives the spectral rigidity $\Delta_3(L)$. For the Poisson case, when the 
eigenvalues are uncorrelated, $\Delta_3 (L) = L/15$, reflecting strong fluctuations 
around the spectral density $\rho(\lambda)$. On the other hand, for the GOE case, 
$\Delta_3(L)$ depends {\it logarithmically} on $L$, i.e.,
\begin{equation}
\Delta_3(L) \sim \frac{1}{\pi^2} \ln L.
\label{del3_GOE}
\end{equation}

\begin{figure}[t]
\centerline{
\includegraphics[height=7cm,width=8cm]{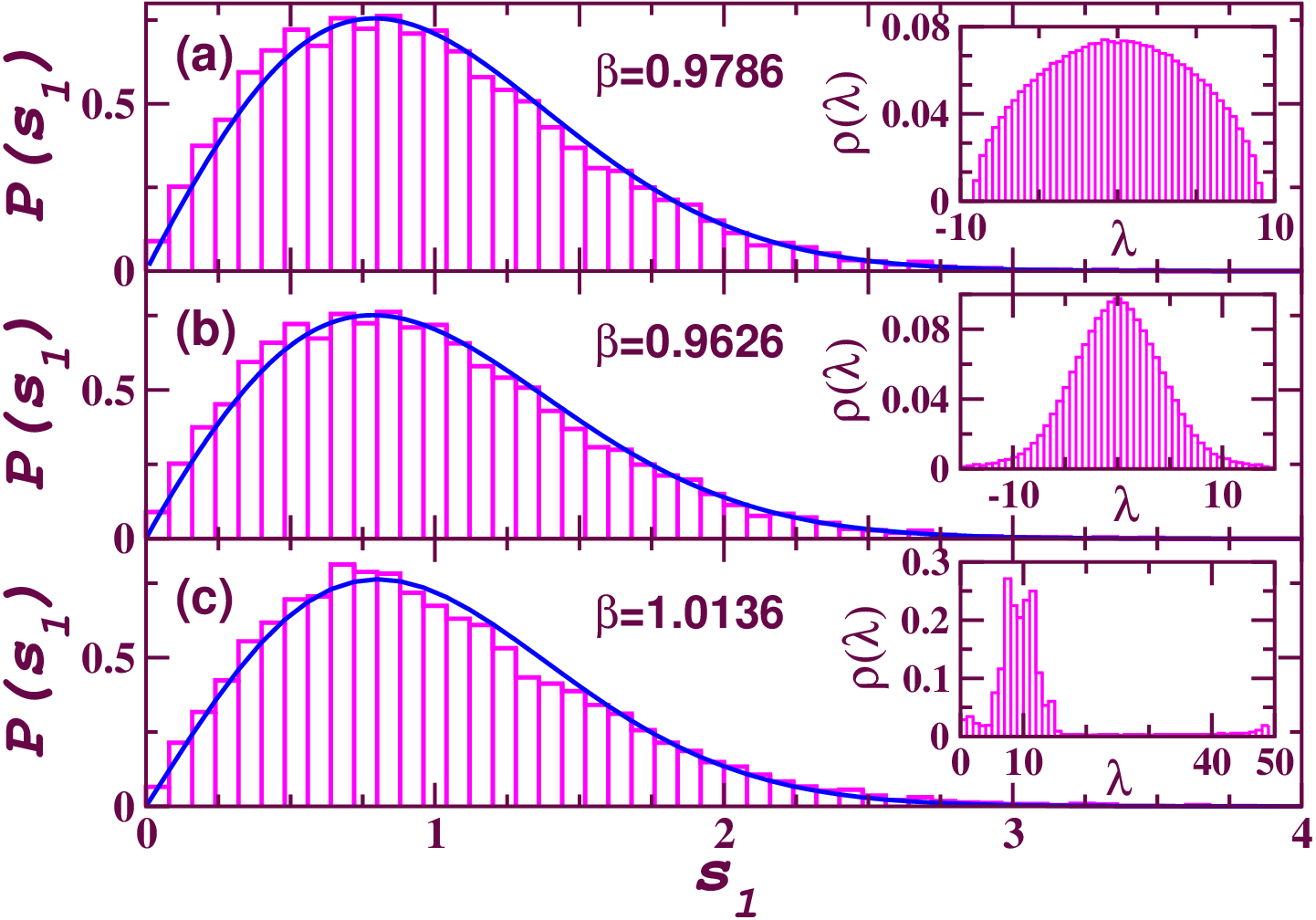}}
\caption{(Color online) Nearest-neighbor spacings distribution (NNSD) $P(s_1)$ of 
the adjacency matrices of different networks [(a) random network, (b) scale-free 
network, and (c) small-world network]. All follow GOE statistics. The histograms are 
numerical results and the solid lines represent fitted Brody distribution (Eq.(3)). 
All networks have $N=2000$ nodes and an average degree $k = 20$ per node. Figures 
are plotted for average over 10 random realizations of the networks. Insets show 
respective spectral densities $\rho(\lambda)$.}
\label{nnsd}
\end{figure}

\begin{figure}[t]
\centerline{
\includegraphics[height=7cm,width=8cm]{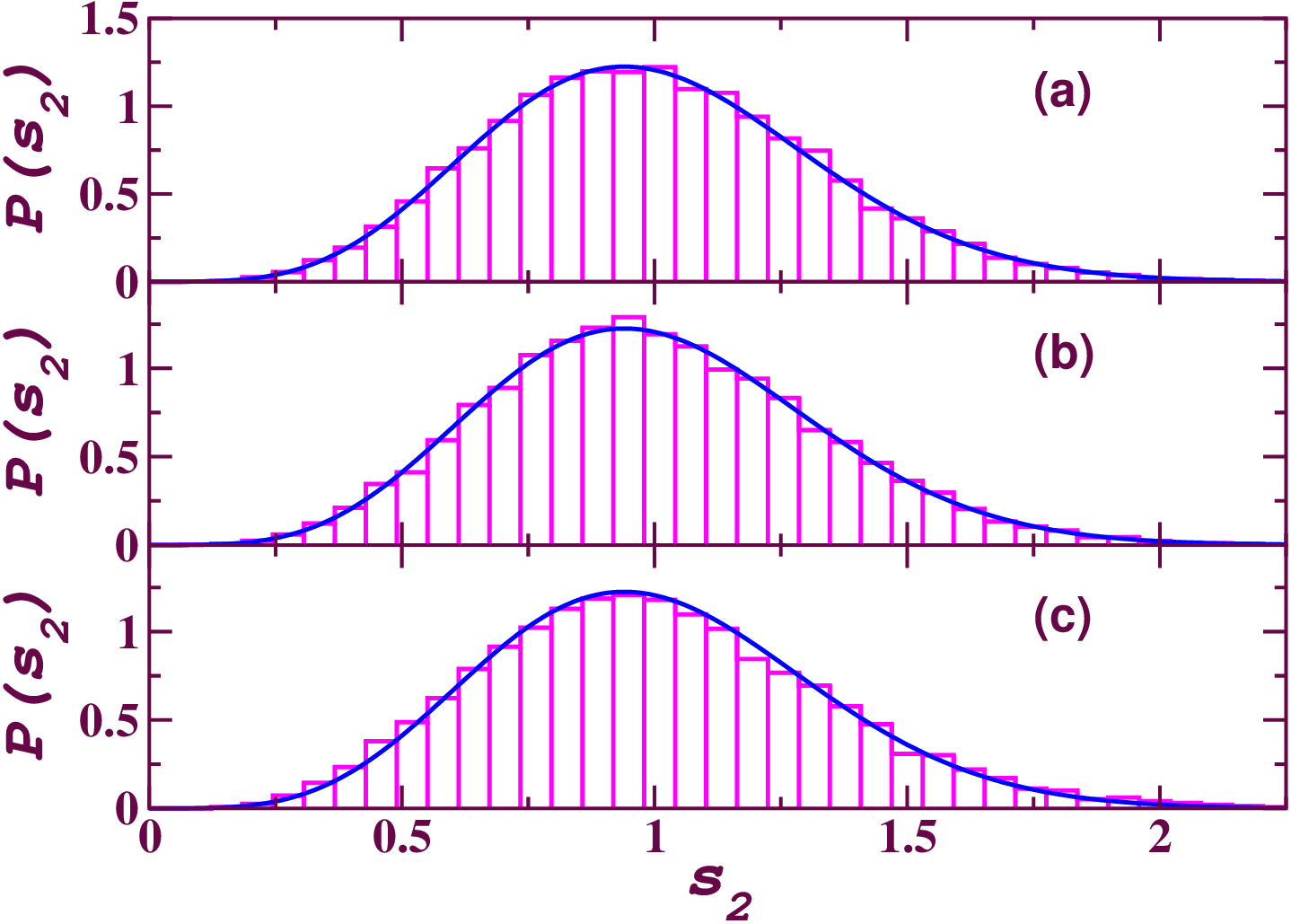}}
\caption{(Color online) Next-nearest-neighbor spacings distribution (nNNSD) $P(s_2)$ 
of the adjacency matrices of different networks [(a) random network, (b) scale-free 
network, and (c) small-world network] is compared with the nearest-neighbor spacings 
distribution (NNSD) of GSE matrices. Figures are plotted for average over 10 
realization of the networks. All networks have $N=2000$ nodes and an average degree 
$k=20$ per node.}
\label{nnnsd}
\end{figure}

\section{Results}
\label{results}

In the following we present results for the ensemble averaged NNSD, nNNSD and 
$\Delta_3$ statistic of random, scale-free and small-world networks.

\begin{figure}[t]
\centerline{
\includegraphics[height=7cm,width=8cm]{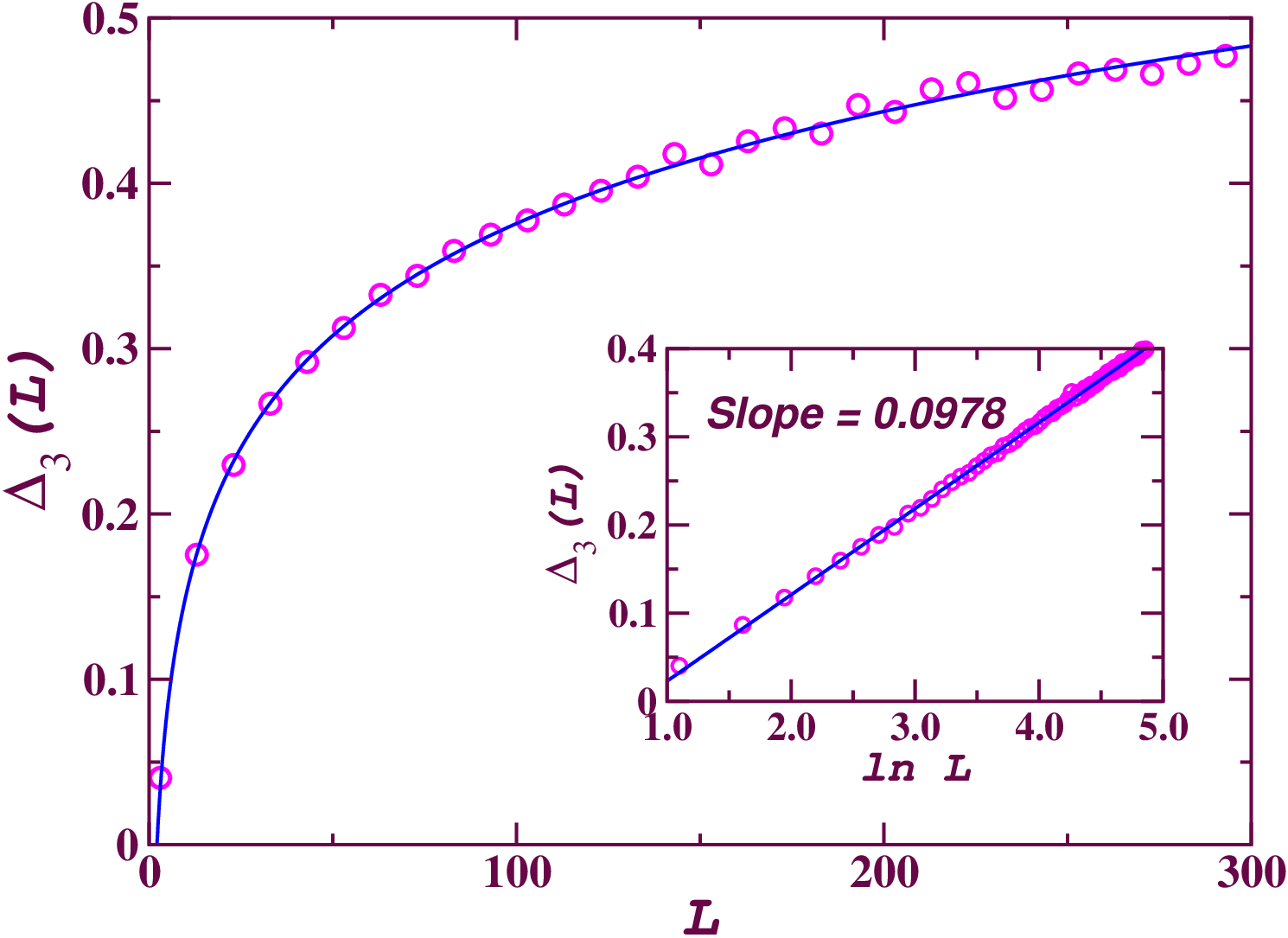}}
\caption{(Color online) $\Delta_3(L)$ statistic for eigenvalues spectra of the 
random network. The circles are numerical results and the solid curve is GOE 
prediction of RMT. Inset shows the $\Delta_3(L)$ in semi-logarithmic scale, in this 
scale it has the slope $0.0978$. Figure is plotted for average over 10 realizations 
of the networks. All networks have $N=2000$ nodes and an average degree $k = 20$ per 
node.}
\label{rand-delta}
\end{figure}

\subsection{Random network}
\label{result_ER}

First we consider an ensemble of random networks generated by using 
Erd\"os-R\'{e}nyi algorithm. Starting with $N=2000$ nodes random connections between 
pairs of nodes are made with probability $p$. The average degree of the graph is $k 
= 2n/N = p(N-1) \sim pN$. There exists a critical probability $p_c(N)$ for which one 
gets a large connected component. The degree distribution of this random graph is a 
binomial distribution $P(k) = C^k_{N-1} p^k (1-p)^{N-1-k}$.  For $p=0.01$, this 
method yields a connected network with average degree $p \times N = 20$. Note that 
for very small value of $p$ one gets several disconnected components.  In this study 
choice of $p$ is high enough to give a connected component typically spanning all 
the nodes.

We calculate the eigenvalues spectrum of network generated according to the above 
algorithm. First the eigenvalues are unfolded by using the technique described in 
Sec. \ref{RMT}. This method yields the eigenvalues with constant spectral density on 
the average. These unfolded eigenvalues are used to calculate NNSD. The same 
procedure is repeated for an ensemble of the networks generated for different random 
realizations. Note that $p$ is always chosen such that algorithm generates a network 
with average degree $k=20$. Fig.\ref{nnsd}(a) plots ensemble average of NNSD. By 
fitting this ensemble averaged NNSD with the Brody formula given in 
Eq.(\ref{eq-brody}) we get an estimation of the Brody parameter $\beta = 0.9786 \sim 
1$.  This value of Brody parameter clearly indicates the GOE behavior of the NNSD 
[Eq.~(\ref{goe})]. Inset of Fig.~\ref{nnsd}(a) shows corresponding spectral density 
which follows well known Wigner's semicircular distribution. The same unfolded 
eigenvalues are used to calculate nNNSD. For this we calculate next nearest neighbor 
spacings as given in Eq.~(\ref{eq-s2}) and plot their distribution in 
Fig.~\ref{nnnsd}(a). It can be seen from the figure that the nNNSD agrees well with 
the NNSD of GSE matrices as given in Eq.~(\ref{spacing_GSE}).

As explained in the introduction that NNSD and nNNSD only tell about the short range 
correlations among the eigenvalues. Therefore, to probe for the long range 
correlations we study $\Delta_3(L)$ statistic of the spectrum of this network. 
$\Delta_3 (L)$ is calculated following Eq.~(\ref{eq-delta3}). Fig.~\ref{rand-delta} 
plots this statistics for the same ensemble as used for the NNSD and nNNSD 
calculations above. It can be seen that $\Delta_3(L)$ statistic agrees very well 
with the RMT prediction, given by Eq.~(\ref{del3_GOE}), upto very large value of 
$L$, i.e., $L \sim 300$. Inset of this figure shows the same in semi-logarithmic 
scale. Here one can see the expected linear behavior of $\Delta_3(L)$ with slope of 
$0.0978$ which is very close to the RMT predicted value $1/\pi^2 \sim 0.1013$ 
[Eq.(\ref{del3_GOE})].

Note that here an ensemble of ten networks of dimension $N=2000$ is considered. 
Statistical properties of eigenvalues spectra of members of this ensemble have very 
small deviations from each other and hence justifying ensemble averaging 
calculations \cite{ergodicity-RMT}. Each individual network in the ensemble follows 
random matrix predictions with very good accuracy, however to make the statistical 
analysis more credible, we present the results for an ensemble of ten networks. Here 
we would like to mention that an ensemble of networks of much smaller dimensions, 
say $N=100$, has been studied as well and it follows GOE predictions of RMT. 
However, for this case, many more realizations are required to get good accuracy.

\subsection{Scale-free network}

Scale-free network is generated by using the model of Barab\'asi and Albert 
\cite{BA}. Starting with a small number, $m_0$ of the nodes, a new node with $m \leq 
m_0$ connections is added at each time step. This new node connects with an already 
existing node $i$ with probability $\pi(k_i) \propto k_i$, where $k_i$ is the degree 
of the node $i$. After $\tau$ time steps the model leads to a network with $N=\tau + 
m_0$ nodes and $m \tau$ connections. This model generates a scale-free network, 
i.e., the probability $P(k)$, that a node has degree $k$ decays as a power law $P(k) 
\sim k^{-\lambda}$, where $\lambda$ is a constant and for the type of probability 
law $\pi(k)$ used here $\lambda = 3$. Other forms for the probability $\pi(k)$ are 
also possible which give different values of $\lambda$. However, the results 
reported here are independent of the value of $\lambda$.

\begin{figure}[t]
\centerline{
\includegraphics[height=7cm,width=8cm]{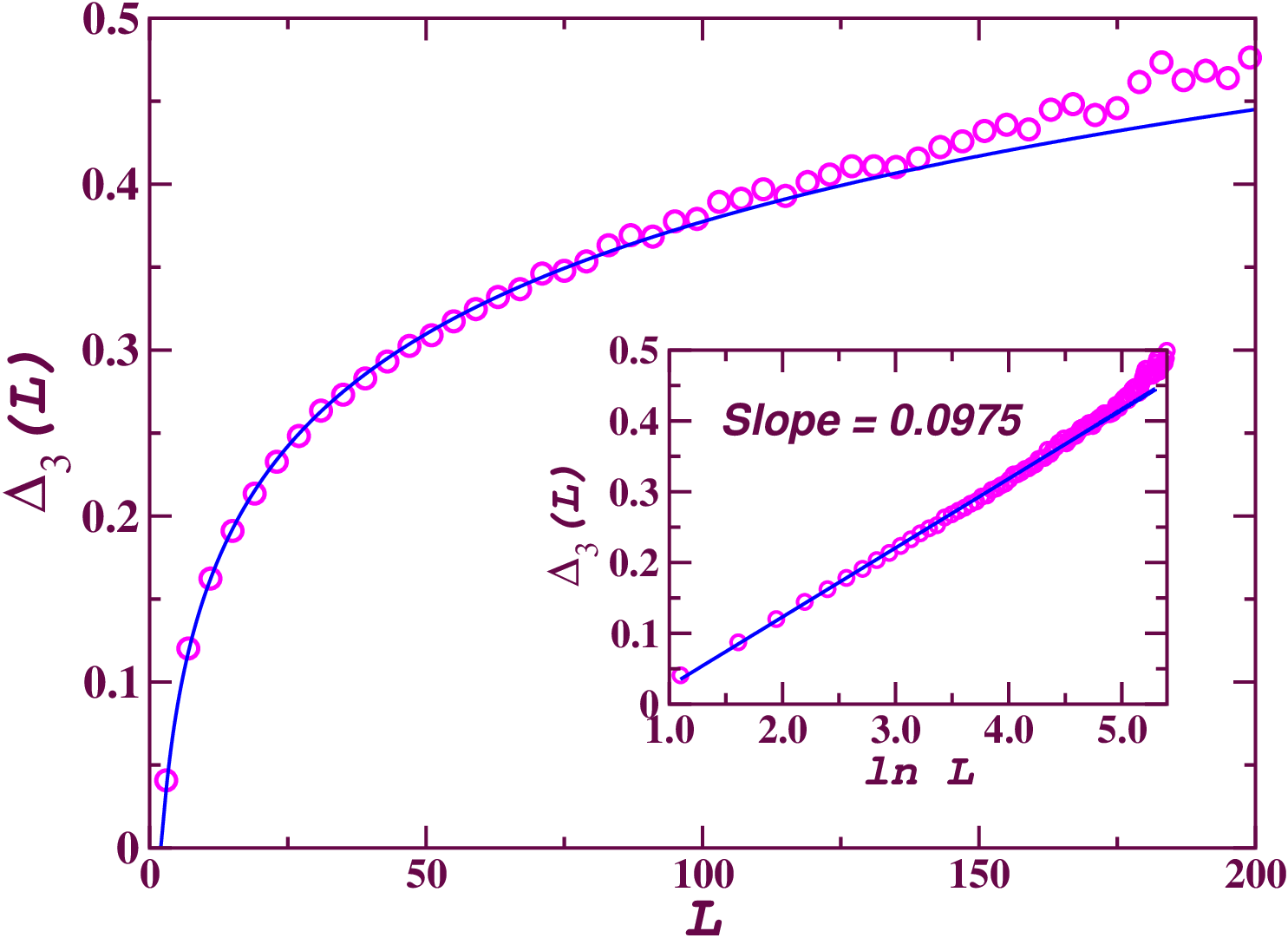}
}
\caption{(Color online) $\Delta_3(L)$ statistic for eigenvalues spectra of the 
scale-free network. The circles are numerical results and the solid curve is the GOE 
prediction of RMT. Inset plots the $\Delta_3(L)$ in semi-logarithmic scale, in this 
scale it has the slope $0.0975$. Figure is plotted for the average over 10 
realizations of the networks. All networks have $N=2000$ nodes and an average degree 
$k = 20$ per node.}
\label{SF-delta}
\end{figure}

Using the above algorithm an ensemble of scale-free networks of size $N=2000$ and 
average degree $k=20$ is generated. To calculate NNSD, nNNSD and $\Delta_3$ for the 
spectra of this ensemble, we follow the same procedure as described in the previous 
section. Fig. \ref{nnsd}(b) shows that the NNSD of scale-free network follows GOE 
with $\beta = 0.9626 \sim 1$. Inset of this figure plots the spectral density of 
this network. Fig.~\ref{nnnsd}(b) shows that the nNNSD of the adjacency matrix of 
this network agrees well with the NNSD of the GSE matrices. Fig.~\ref{SF-delta} 
shows the $\Delta_3(L)$ statistic for the adjacency matrix of scale-free network. 
Here we see that the $\Delta_3(L)$ statistic for the scale-free network agrees very 
well with the RMT prediction for very large $L$, i.e., $L \sim 150$, and deviations 
are seen only after $L=150$. Inset of this figure shows the expected linear behavior 
of $\Delta_3(L)$ in semi-logarithmic scale for $L \lesssim 150$ with the slope of 
$0.0975$, a value very close to the RMT predicted value $1/\pi^2$.

Universality of NNSD and nNNSD for random and scale-free networks seems to give the 
impression that these networks have same amount of randomness but $\Delta_3$ results 
tell that the scale-free network is not as much random as the random network. This 
is obvious from their construction algorithms as well, but $\Delta_3$ statistics is 
capturing this property which is a very important result. The finding also suggests 
that scale-free networks have some specific features that cannot be modeled by RMT. 
It may be noted that one can generate scale-free networks by using other algorithms 
as well \cite{kim,mendes}, for these networks also spacing distributions and 
spectral rigidity results will have qualitatively similar behaviors, except that the 
range of agreement of $L$ with the random matrix prediction would depend upon the 
amount of randomness in the networks.

\subsection{Small-world network}

Small-world networks are constructed using the following algorithm of Watts and 
Strogatz \cite{SW}. Starting with an one-dimensional ring lattice of $N$ nodes in 
which every node is connected to its $k/2$ nearest neighbors, we randomly rewire 
each connection of the lattice with the probability $p$ such that self-connections 
and multiple connections are excluded. Thus $p=0$ gives a regular network and $p=1$ 
gives a completely random network. The typical small-world behavior is observed 
around $p=0.005$ \cite{pap1}. For $N=2000$ and average degree $k=20$, an ensemble of 
{\it ten} different realizations of the network are generated.

\begin{figure}[t]
\centerline{
\includegraphics[height=7cm,width=8cm]{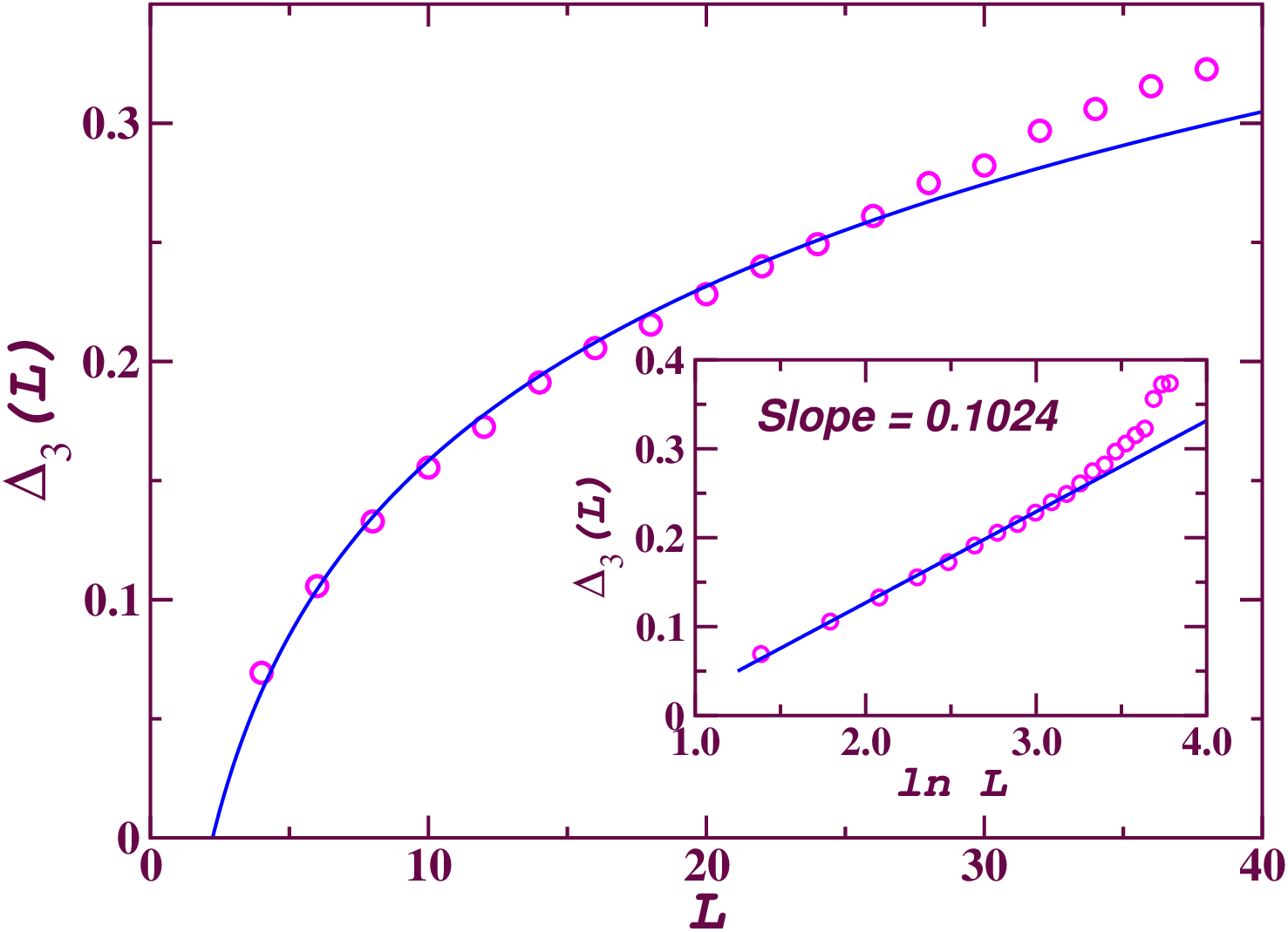}
}
\caption{(Color online) $\Delta_3(L)$ statistic for the eigenvalues spectra of the 
small-world network. The circles are numerical results and the solid curve is GOE 
prediction of RMT. Inset plots the $\Delta_3(L)$ upto $L = 30$ in semi-logarithmic 
scale, in this scale $\Delta_3(L)$ has the slope $0.1024$. Figure is plotted for the 
average over 10 realizations of the networks. All networks have $N=2000$ nodes and 
an average degree $k = 20$ per node.}
\label{SW-delta}
\end{figure}

Again the same procedure as described in Sec.~\ref{result_ER} has been used to 
calculate NNSD, nNNSD and $\Delta_3$ for the spectra. Fig.\ref{nnsd}(c) shows that 
the NNSD of this network again follows GOE statistics with $\beta$ very close to 
$1$, i.e., $\beta = 1.0136$. The inset shows that the corresponding spectral density 
is complicated with several peaks. One peak is always at $\lambda = 0$.  The exact 
positions of other peaks may vary but overall form of spectral density remains 
similar. Fig.~\ref{nnnsd}(c) plots the nNNSD of adjacency matrix of small-world 
network. It can be seen that the nNNSD agrees well with the NNSD of GSE matrices.  
Fig.~\ref{SW-delta} shows the $\Delta_3(L)$ statistic for the spectrum of adjacency 
matrix corresponding to the small-world network with $p = 0.005$. Inset of this 
figure shows the expected linear behavior of $\Delta_3(L)$ in semi-logarithmic scale 
for $L \lesssim 30$ with slope of $0.1024$, a value very close to the RMT predicted 
value $1/\pi^2$. It can be seen here that $\Delta_3(L)$ statistics for the 
small-world network agrees very well with the RMT prediction for sufficiently large 
$L$, i.e., $L \sim 30$, but much less than the same for random and scale-free 
networks, implying that besides randomness, small-world network has specific 
features also. This again suggests that the behavior of $\Delta_3$ statistics can be 
used to understand the amount of randomness in the networks. More specifically 
deviation from the GOE predicted behavior corresponds to the system specific 
features in the network.

Note that in this paper, results for networks with the average degree 20 are 
presented. We have studied sparser ($\langle k \rangle < 20$) and denser networks 
($\langle k \rangle > 20$ to $\langle k \rangle \sim N$) as well. Same universal 
behavior are found for these networks as far as there exists a certain amount of 
randomness, i.e., presence of some {\it minimal} random connections among the nodes. 
There exists problems with very sparse networks as of average degree {\it two} and 
very dense networks as of degree $\sim N$. For sparse networks ($\langle k \rangle 
\sim 2$) sometimes one can get several degeneracies in eigenvalues \cite{sparseRM}. 
In this case, one has to first get rid of the degeneracies to conclude anything 
under the RMT framework \cite{note}.

Similarly, for dense networks, universal spacing distribution are observed till very 
large value of average degree. For $\langle k \rangle \sim N$, largest eigenvalue 
has very high value than the rest of $N-1$ eigenvalues which are very close to each 
other, becoming equal in the limiting case of all to all connections $\langle k 
\rangle = N-1$. For example, random networks with $p=0.95$ (which means that the 
network has $95\%$ of maximum possible connections), also follow RMT predictions of 
universal spacing distributions till very large scale. As the number of connections 
are increased further one starts getting degenerate eigenvalues and for $p=0.999$ 
high degeneracies at various values (such as $\lambda= -1, 0 ,1$) are observed 
keeping it trivially out from the RMT studies.

\section{Discussions}
\label{summary}

We use RMT to study complex networks and show that in spite of spectral densities of 
the adjacency matrices being different for different networks, their eigenvalue 
fluctuations are same and follow the GOE statistics of RMT. We attribute this 
universality to the existence of {\it minimal amount of randomness} in all these 
networks and show that randomness in the network connections can be quantified by 
the Brody parameter. In addition to the NNSD, we present the results of nNNSD and 
spectral rigidity via $\Delta_3$-statistic of RMT. The nNNSD of the eigenvalues of 
these model networks are identical to the NNSD of GSE matrices which again agrees 
with RMT prediction given in Eq.~(\ref{spacing_GSE}). NNSD and nNNSD suggest that 
there exists short range correlations among the eigenvalues. Spectral rigidity test 
shows that the $\Delta_3(L)$ statistics follows random matrix predicted linear 
behavior in semi-logarithmic plot for sufficiently large scale $L$ with slope being 
$\sim 1/\pi^2$ [see, Eq.(\ref{del3_GOE})], suggesting long-range correlations among 
the eigenvalues. Above findings show that statistics of the bulk of eigenvalues of 
the model networks is consistent with those of a real symmetric random matrix and 
deviation from this could be understood as a system dependent part.

Universal GOE behavior of NNSD and nNNSD tell that the networks are sufficiently 
random, or there exists {\it minimal amount of randomness} required to introduce the 
correlations among the neighboring eigenvalues. The $\Delta_3$ analysis seems to 
characterize the level of randomness in networks depending on the range of 
correlations among eigenvalues. $\Delta_3$ analysis of the random network follows 
RMT prediction for very long range of $L$, which is not very surprising as random 
network follows RMT at each level starting from semi-circular density distributions. 
However interestingly scale-free and small-world networks also follow RMT for 
sufficiently large value of $L$. Beyond this value of $L$, deviation in the spectral 
rigidity is seen, indicating a possible breakdown of the universality. This is quite 
understandable as small-world network is highly clustered and scale-free network 
also has specific features like hubs, so it is natural that they are not as random 
as the random network. But it is interesting to realize that $\Delta_3$ statistics 
rightly captures this information. Moreover, small-world network is generated 
exactly at small-world transition by using Watts and Strogatz algorithm which yields 
network with very high clustering coefficient and very less number of random 
connections. Results presented in this paper show that these very small number of 
random connections make network {\it sufficiently} random to introduce the 
correlations among the eigenvalues for the sufficiently long range.

According to the many recent studies, randomness in connections is one of the most 
important and desirable ingredients for the proper functionality or the efficient 
performance of systems having underlying network structures. For instance, 
information processing in brain is considered to be because of random connections 
among different modular structures \cite{face}. We feel that the role of random 
connections, and behavior and evolution of such systems can be studied better under 
the RMT framework. Also this RMT approach may be used to detect the connections most 
responsible to increase the {\it complexity} of networks. For example effect of 
oxygen molecule on biochemical network of a metabolic system is recently studied and 
is shown to increase the complexity of the system leading to a major transition in 
the history of life \cite{oxygen-complex}.

In summary, we use RMT to analyze spectra of complex networks and show that these networks 
follow universal GOE statistics. These results tell that 
random matrix theory, a very well developed branch of Physics, can be applied to the 
complex networks studies. So far we have only concentrated on the model networks 
studied vastly in the recent literature, providing a basis to the random matrix 
analysis of networks. Future investigations would involve studies of real-world 
networks \cite{EV}.

\acknowledgments
We acknowledge an anonymous referee for constructive suggestions, as well bringing an important 
reference \cite{bauer} to our notice. SJ acknowledges Max-Planck Institute for Mathematics 
in the Sciences, Leipzig, where part of this work was done.

\end{document}